\documentclass[letterpaper,12pt]{article}
\usepackage{fullpage}
\usepackage{amsmath}
\usepackage{amsfonts}
\usepackage{amssymb}
\usepackage{subfigure}
\usepackage{graphicx}

\begin{document}

\begin{titlepage}
\author{
Andrew Keenan Richardson\hspace{.3in}
Cole Arthur Brown
}
\title{A Critique of ``Solving the P/NP Problem Under Intrinsic Uncertainty''}
\maketitle
\thispagestyle{empty}

\begin{abstract}
Although whether P equals NP is an important, open problem in computer science, and although Jaeger's 2008 \cite{jaeger08} paper, ``Solving the P/NP Problem Under Intrinsic Uncertainty'' presents an attempt at tackling the problem by discussing the possibility that all computation is uncertain to some degree, there are a number of logical oversights present in that paper which preclude it from serious consideration toward having resolved P-versus-NP.  There are several differences between the model of computation presented in Jaeger's paper \cite{jaeger08} and the standard model, as well as several bold assumptions that are not well supported in Jaeger's paper \cite{jaeger08} or in the literature.  In addition, we find several omissions of rigorous proof that ultimately weaken this paper to a point where it cannot be considered a candidate solution to the P-versus-NP problem.
\end{abstract}
\end{titlepage}

\section{Overview}
Jaeger \cite{jaeger08} presents a paper which attempts to show that P is not equal to NP.  This follows from a novel model of computation which has intrinsic uncertainty in its computation of all problems.  However, we will show that this model cannot be used to solve the same class of decisions that a Turing machine can.

\subsection{Summary of the Paper in Question}
Jaeger \cite{jaeger08} in his paper: Solving the P/NP Problem Under Intrinsic Uncertainty, attempts to resolve P-versus-NP through unique, unconventional means, by sketching a computer science analogue to the Heisenberg Uncertainty Principle.  He begins by establishing a supposedly Turing equivalent computing machine largely based off of the Turing machine that contains a tape of binary storage cells of any length, infinite or finite.  Rather than assuming the machine is preprogrammed to accept certain inputs, the model described has a single tape in which both the input and the program code are randomly placed.  This allows him to introduce the uncertainty that the paper is based on.  He argues that since multiple, unique programs can perform the same actions at different speeds, program size should be considered when analyzing program complexity.  He continues that his concept of “intrinsic uncertainty” implies that it is impossible to distinguish whether, given a segment of the input, a part of this segment is program code or input.

Jaeger \cite{jaeger08} likens the relationship of the parts of this segment to the relationship described in Heisenberg’s Uncertainty Principle -- that you can not determine whether a part is code or input, rather you can compute the answer given both options and then compute the “certainty” that each computation is correct.  There is a problem with this, he explains, because according to his model every computation done on the machine has uncertainty, including his certainty calculations.  To solve this problem, he proposes what he calls “self-computation”, or applying his Turing-machine-based machine to itself to compute a confidence value for its output, thus increasing the confidence value of its output.  As a result of this self-computation he claims that the confidence value of a computation is directly proportional to the ratio of code length to input length.  Through use of the sigmoid function, he claims one can calculate the entropy, or the probability that this confidence value is smaller than, larger than or equal to 0.5.

To relate this all to the P/NP problem, Jaeger \cite{jaeger08} constructs three lemmas.  His first lemma says that a Turing machine simulating a computable decision in NP to an arbitrary precision is itself computable regardless of the precision.   Lemma three states that this Turing machine is also in NP for some given uncertainties.  Lemma two allows him to state an uncertainty threshold  for which the Turing machine is not in P, thus proving that there exists some  function in NP but not in P.

\subsection{P = NP Problem}
The problem known as P=NP is arguably the biggest open problem in computer science \cite{sipser92}.  The Clay Mathematics Institute has offered a \$1,000,000 prize for anyone who provides a proof one way or another.  This incentive has attracted quite a few attempts by amateurs as well as professionals, since the problem at first seems deceptively simple.  The basic question asked by this problem is this: if an answer to a yes or no problem can be verified in polynomial time, can the answer also be computed in polynomial time?  The answer is commonly assumed to be `no', although no proof has yet emerged.  

One way to prove that P is not equal to NP is to give a counter example of a problem that is in NP but proveably not in P.  This is more difficult than it sounds because it is difficult to prove that there are no algorithms in P which compute the answer.  Jaeger's paper \cite{jaeger08} attempts to prove that there is some inrinsic information necessary for having the answer to a problem, and that this intrinsic information can only be computed in nonpolynomial time.

\subsection{Turing Machines}
Jaeger 2008 \cite{jaeger08} uses the model of a Turing machine, a theoretical model of computing yes or no answers.  The model of a Turing machine has several parts \cite{sipser05}: a finite set of states including the initial state and a subset of final or accepting states, a finite set of tape alphabet symbols composed of the blank symbol and a set of input symbols, and a transition function which computes the new state, the alphabet symbol to be printed, and the direction to move (right or left), based on the current state and the current tape symbol.  Informally there are two parts in a Turing machine: a finite set of states in which the current state is arrived at deterministically, and an infinite tape which may start with a finite number of non-blank symbols printed on it.  

\section{Challenging the Arguments Made}
We will show that the model presented in this paper is flawed because it is not equivalent to a Turing machine.  Moreover, the self-computation algorithm presented here has several major shortcomings which prevent it from resolving that uncertainty.

\subsection{Theoretical Differences with Turing Machines}
The most fundamental flaw to the argument presented in the paper under consideration is that rather than using a theoretically established model like a Turing machine, it presents a model in which the ``code" of the model, presumably corresponding to the set of states and the transition function, is indistinguishable from the ``input" to the code, presumably corresponding to the initial content of the tape.  Although there are numerous very obvious theoretical and practical ways of distinguishing ``code" from ``input", they are not clearly marked in this model, causing the intrinsic uncertainty upon which the core argument is based.

``Let us assume in the following that we have a Turing-complete machine architecture...", reads paragraph 2, pg 5.  The paper seems to rely on this model being Turing-complete, a term which is never clearly defined but which we assume to mean Turing equilavent.  However, this is clearly not the case, since the elements of the 7-tuple defining a Turing machine are clearly identifiable, and the same can be said for the initial state of the tape.  This method of arbitrarily introducing uncertainty to the computation of a Turing machine does not produce a Turing equivalent model, as it cannot be used to simulate a Turing machine, nor can it compute the same class of decisions that a Turing machine can.  Because the output of this model is always uncertain, there are no problems in NP that can be reduced to it.  It is similar to a model that flips a coin and upon seeing a heads, returns an arbitrary answer. 

\subsection{Lack of Rigor in Analysis}
The definition of the machine the proof revolves around is seemingly flawed.  The machine which is supposed to be equivalent to a Turing machine in computational power detailed has a tape that contains both its program code and input, with both being randomly distributed throughout the tape.  The machine reads the tape which is partitioned into two subsets, $S_1$ and $S_2$.  We are lead to assume that these subsets form the direct sum of the original input, but this is never verified.  As a result, one could infer that perhaps there is some intersection between $S_1$ and $S_2$.  This is an example of the unrigorous definitions throughout the paper.  These subsets are never fully defined - no methods for finding them are ever given.  The machine evaluates the tape by computing the results as if $S_1$ were the code and $S_2$ the input and vice-versa, then calculates confidence values for both of those computations to help decide which configuration is correct.  This has some problems, namely that the method for partitioning the input into two distinct blocks is non-existent.  On page 6, the partition is described as being N bits long, but no further information is given.  Given the fact that the input and code are randomly intermixed, one could go as far as to say that its impossible to partition the input with any precision - both the code and input string are represented in binary and are thus indistinguishable.  This means that not only would the machine have to evaluate for all possible $S_1$-$S_2$ dichotomies, but also for all possible partitions of the input.  This additional factor is left completely unaccounted for in the remainder of the paper.

There is another error on page 6 in the uncertainty section.  The claim that, ''we can accept the interpretation whose program code encompasses the larger number of bits as more likely," is  never justified.  Such a statement is invalid - in many cases, for instance any program dealing with databases, the size of the input vastly exceeds the size of the program code.

One mistake that appears quite often in Jaeger 2008 \cite{jaeger08} is the constant changing of terms.  In section 2, page 5, the machine used throughout the paper is described as a ''Turing-complete" machine based on the Turing machine.  Just two pages later on page 7, the same machine is referred to as a Turing machine.  This inconsistency is detrimental to the reader's understanding as well as the validity of the paper itself.  The paper also mentions an ''outside program" that performs the interpreting and execution aspects of the previously defined computing machine.  While the existence of such a ''program" is provable, the author has omitted any kind of proof that such a program exists.

\subsection{The Self Computation Algorithm}
The bulk of this article describes an algorithm to be used for calculation of certainty measurements to be reported alongside the actual findings of a Turing machine.  The method of self-computation proposed is hugely complex, riddled with nonstandard notation and very confusing; the authors of this paper were unable to make much sense of it.  Self-computation, as described in the paper, involves executing the uncertain Turing machine on a copy of a similar machine running arbitrary ''code".  The chance that the machine could execute a copy of itself and gain higher precision is unlikely and left completely unproven.  The insinuation is that as you run the machine on itself, the bitlength of the ''code" portion gets longer and longer, increasing the ratio of code to data, which is directly proportional to certainty.

While this method may appear useful for calculating uncertainty, it is unlikely that this will work because it assumes, among other things, that bit length is a useful metric for telling code and data apart, that this is a necessary determination in a Turing machine, that Turing equivalent models can report a certainty measurement, and that no other algorithms exist for computing this function.  As we have shown these things to be untrue, the internal merit of the self computation algorithm is irrelevant.

\subsection{There is No Proof of Nonpolynomiality for Certainty Calculation}
One problem with the argumunt presented in Jaeger 2008 \cite{jaeger08} (Lemmas 1 and 3) is that it is nowhere proven that the method of self-computation presented in there is the only method of determining the certainty of a result.  Even granting that there is a problem of determining certainty of answers produced by Turing machines, and given that this method of self-computation is a reasonable way of determining certainty, and given that a certainty threshold can be set such that computing certainty by this method cannot be done in polynomial time, all doubtful claims, nothing in Jaeger's paper \cite{jaeger08} attempts to prove that there is no other method of computing this certainty measurement.  It could very well be that there is a polynomial time algorithm for computing this measurement that the author simply did not think of.  

Indeed, it is unclear from the paper under consideration why it should be possible to set a threshold ''dynamically" so that it requires NP time to reach that level of certainty.  Even using the algorithm outlined there, it is not clearly explained why this would require exponential time rather than polynomial time.  This trick of requiring that an exponential amount of computation go into producing the answer is only possible as a use of the certainty value which is produced by the modified Turing machine.  As the output of a standard Turing machine is binary, this type of trick, which requires using the extra non-binary output, would not normally be possible.  It is as if one constructed an algorithm to run in exponential time by requiring that it print an exponential number of characters.

\section{Conclusion}
Overall, we have shown several things which should each individually render the argument described in Jaeger's paper \cite{jaeger08} impotent.  We have shown that a model of computation that requires a measure of intrinsic uncertainty cannot be reduceable to a Turing machine nor can it reliably compute any NP-complete problem.  We have also shown that the algorithm proposed for calculating uncertainty relies on faulty or unproven assumptions.  Finally, we note that while one algorithm is presented here, which may very well have an exponential runtime, there is no attempt to prove that there cannot be a polynomial-time algorithm to compute the same.  Any of these things strikes a fatal blow to the argument outlined by Jaeger \cite{jaeger08}.  As such, we do not find this compelling support for P not equal to NP.

\section{Acknowledgements}
This work was done as a project in the Spring 2009 CSC 200H course at the University of Rochester.  We thank the professor, Lane A. Hemaspaandra, and the TA, Adam Sadilek, for their comments and advice.  Any opinions, errors, or omissions are the sole responsibility of the authors.  

\bibliographystyle{plain}
\bibliography{pnp3}

\end{document}